\documentclass[%
 reprint,
 amsmath,amssymb,
 aps,
]{revtex4-1}

\usepackage{graphicx}
\usepackage{dcolumn}
\usepackage{bm}

\begin{document}

\preprint{APS/123-QED}

\title{Towards Measuring Vacuum Polarization of Quantum Electrodynamics   with Superconducting Junctions}
\thanks{The Proposal}%

\author{Ali Izadi Rad}

\collaboration{ 
University of British Columbia, Vancouver, Canada}



\begin{abstract}
In this proposal,  we present an experimental setup based on superconducting circuits and Josephson junctions to explore the modification of Josephson coefficient in the presence of external magnetic field due to vacuum polarization of quantum electrodynamics. This robust experiment can be considered as one of the few possible chances to observe the fine quantum field theory corrections in the low energy regimes in condensed matter systems. It can also be a new check for the universality of Josephson constant which is important in metrology. We will expect the signal to noise ratio of the read-out signal to increases quadratically by running time of the experiment. This characteristic of the output signal the will  guarantee the feasibility of measurements with desired precision

\end{abstract}

\maketitle


\section{Introduction}

Superconductivity is one of the most interesting macroscopic quantum phenomenon which provides a window to observe quantum mechanical behaviors in a classical scales. As an  example, we can observe the tunneling effect by considering two piece of superconductor located on two sides of one insulator as a barrier. If we apply fixed voltage across them, the tunneling of wave functions of Cooper pairs  in each sides provides a current of charge across the barrier which is known as  AC Josephson  effect. It has been observed \citep{100} applying constant voltage $V$ across the junction creates alternating current with frequency $f$, which is linearly related to V,
\begin{equation}
 \nu=K_j V=\frac{2e}{h}V.   
\end{equation}

Josephson effect and Quantum Hall effect are two well know phenomena in condensed matter physics that  are robust  against any perturbation due to their gauge invariance property.
It is now generally accepted that Josephson Constant, $K_j=484\text{ GHz/mV}$, is invariant under the design of junction with very high accuracy \cite{5}\cite{6}. Josephson constant in the recent decades has been considered as a precise constant for metrology and a standard way to measuring voltages\cite{10}.

Application of superconducting  circuits in high precise measurement has been developed for decades and today detection of very small electromagnetics flux by superconducting quantum interference device (SQUID) is a routine  work and counting the number of charged particle leaving a typical  superconducting island is under control \cite{1000}. Beside the normal application of superconducting devices for measurement of small Electromagnetic flux and it's application in industry and medicine, gauge invariance properties  of some quantities in superconductivity let us to measure some fundamental parameters. The most accurate measurement of Planck constant is based on the  Josephson constant and quantum Hall coefficient( is called Von Klitizing constant: $R_k=h/e^2$).  Historically the Josephson junction played the role on some new measurement of quantum electrodynamic (QED) constants with high precision \cite{9} such as the  measurement of the fine structure constant, $\alpha=\frac{e^2}{4\pi \epsilon_0 \hbar c} $ in the scale of $\sim 500\text{Kev}$\cite{9}  \footnote{$\alpha=1/137.03599976(50)$.}

 In this proposal we introduce new set up to measure vacuum polarization of quantum electrodynamics and it's affect on Josephson Coefficient. We expect the electric charge of electron in Josephson constant becomes the function of external magnetic filed due to QED corrections:

 \begin{equation*}
     e(B)=e_0[1+\epsilon B^2+ \mathcal{O}(\epsilon^2)].
 \end{equation*}
 
  In the condensed matter physics  electrodynamics interaction play the mains role and in the context of quantum field theory photon is propagator if force and charge of electron describes coupling of interaction. According to quantum field theory, the propagator which caring the force between two interacting particle, is not necessarily simple and it is possible to have many virtual loops in Feynman diagram of process. In the quantum electrodynamic (QED) interactions, the virtual photon which connect two charge fermion particle such as two electron can has a virtual fermion loop. If we estimate the whole loop we obtain  renormalized propagator of photon. This phenomena called vacuum polarization or photon self-energy.

Introducing vacuum polarization in quantum electrodynamics causes nonlinear phenomena. As a result of pair creation, the physical vacuum becomes a medium with dielectric properties and therefore the electric charge of electron make a dependency  on field strength and characteristic momentum transfer. 

 Finding the trace of vacuum polarization inn the condensed matter systems is quite hard and amount of modification in the  most cases is so tiny that it cannot be seen from background noises. The relativistic nature of quantum electrodynamics implies this effect becomes important in very strong field and high energy process.

 Quantum hall effect and Ac Josephson effect are two phenomena that are robust against any external perturbation because of the gauge invariance property and they are on of few candidates to . In quantum hall effect the Johnoson-Nyquist noise limits  us to see fine corrections and it seems there is no quite chance to see novel effects. In the other hands precise experiments are available by superconducting materials. In this paper show how this idea can works.

\section{Effective Electric Charge  in presence of External Magnetic Field}

In quantum electrodynamics, photon as a gauge boson is a  carrier of the electromagnetic force. A simple interaction of two electrons  in theory of QED can be illustrated via Feynman diagrams similar to  in Fig.1. According to quantum field theory vacuum between interacting particles are not merely empty. There is a possibility that photon  creates a short-lived virtual particle-anti particle, Fig.2. The vacuum of quantum electrodynamics is affected by the whole contribution of these pair production and annihilation, and this phenomena which is called vacuum polarization or self-energy of a photon can influence the distribution of charges and effective forces that particles are feeling. It can be compared with polarization in dielectrics.


\begin{center}
\begin{figure}[h]
  \includegraphics[height=26mm]{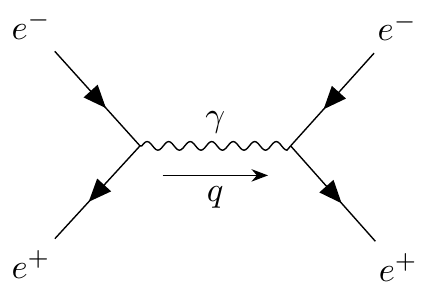}
  \caption{
  Feynman diagram to illustrating the simplest interaction between two charged particle in theory of quantum electrodynamics. The virtual photon (wavy line) carries electromagnetic force between two charge particle( here two electrons).
 }
  \end{figure}
  \end{center}
  
  \begin{center}
\begin{figure}[h]
  \includegraphics[height=15mm]{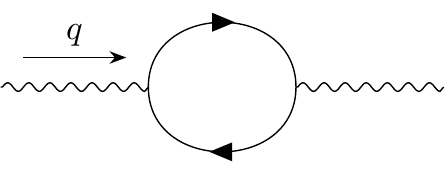}
  \caption{
  First order corrections to interaction forces between two charged particle obtains by considering one loop ( fermion- anti fermion creation and annihilation) in photon's  propagation path. In this paper, we want to study affects of this loops and their correction in superconducting circuits.
 }
  \end{figure}
  \end{center}

If we denote the momentum of virtual photon in Feynman diagram by $q$, then 
 the photon's propagator for diagram in  Fig.1 will given by  $iG_{\mu\nu}=-\frac{g_{\mu\nu}}{q^2}$.The modified and  dressed propagator up to  be  one loop level in Feynman gauge given by:
\begin{equation}
iG_{\mu\nu}=-i\frac{g_{\mu\nu}}{q^2}+\frac{-ig_{\mu\rho}}{q^2}[i(q^2 g^{\rho\sigma}-q^{\rho}q^{\sigma})i{\Pi}(q^2)]\frac{-ig_{\sigma\nu}}{q^2}
\end{equation}
where ${\Pi}(q^2)$ is a mathematical regularization function and incoming electrons with higher momentum leads to higher order correction.


It has been long studied that the vacuum of quantum electrodynamics  in presence of external electromagnetic field would be modified\cite{1}\cite{2}. Calculations show \cite{6} the modified vacuum polarization tensor due to magnetic field in low energy regime, $q \rightarrow 0$  yeilds to 
\begin{eqnarray}
\delta {\Pi}_{\mu\nu}(q)&& = -\frac{\alpha}{45 \pi}(\frac{\hbar eB}{c^2 m^2})[2(g_{\mu\nu} q^2-q_{\mu}q_{\nu}) \\ \nonumber
&& -7(g_{\mu\nu}q^2-q_{\mu}q_{\nu})_{\parallel} +4(g_{\mu}{\nu} q^2-q_{\mu}q_{\nu})_{\perp}]
\end{eqnarray} 
The renormalization of electrical charge is relate to $\delta e \propto {\Pi}_{\mu\nu}(q)/q^2|{q^2 \rightarrow 0} $. This correction can be easily find from modified Coulomb potential \cite{6}:

\begin{eqnarray}
V_C(\mathbf{r})&&=e^2 \int \frac{d^3\mathbf{q}}{(2\pi)^3}\frac{e^{-i\mathbf{q\cdot r}}}{{\mathbf{q}}^2}(1-\frac{\delta {\Pi}_{00}(q)}{q^2})\\  \nonumber
&&= \frac{\alpha}{|\mathbf{r}|}[1+\frac{\alpha}{\pi}(\frac{\hbar eB}{c^2 m^2})^2(\frac{2}{45}-\frac{7}{90}\sin \theta^2)],
\end{eqnarray}

The above equation implies effective electric charge in presence of external fix magnetic field is
\begin{equation}
{e}_{\text{eff}}=e_0[1+\frac{1}{45}\frac{\alpha}{\pi}(\frac{\hbar eB}{c^2m^2})^2 +\mathcal{O}(\alpha^2)]
\end{equation}
as we can see from above equation, the effective electrical charge will change in presence of electromagnetic field from $e$ to $e_{\text{eff}}$. To see affect of this modification on Josephson coefficient we just need to recall that  phase difference between two random  point $x_1$ and $x_2$ in presence of magnetic potential $ A^{\mu}$ given by \cite{9}
\begin{eqnarray}
\Delta \phi&&= \Delta \phi (B,A_\mu=0)-\frac{1}{h}\int_{x_1}^{x_2}2e_{\text{eff}} \mathbf{A} \cdot d \mathbf{x}\\ \nonumber
&&=\Delta \phi (B,A_\mu=0)-\frac{2e_{\text{eff}}}{h}\Delta V,
\end{eqnarray}
therefore if we define the bare $K_J$ as $K_J:=2e/h$, then in general $K_J$ will change to
\begin{eqnarray}\label{kj}
K_J(B)&&=K_J[1+\frac{1}{45}\frac{\alpha}{\pi}(\frac{\hbar eB}{c^2m^2})^2] \\ \nonumber
&&=K_J[1+\frac{\alpha}{45 \pi }(\frac{B}{B_0})^2]
\end{eqnarray}
where $B_0 \sim 10^9 T$ is a critical magnetic field that QED correction can not be ignored at all. 

Despite the very small correction we get for Josephson constant, and by considering that applied magnetic field to junctions are limited by other phenomena,  We will show in the following sections that this correction and effect is measurable due to robustness of of superconducting circuits. 

\section{Measuring the Vacuum Polarization}

\begin{center}
\begin{figure}[h]
  \includegraphics[height=95mm]{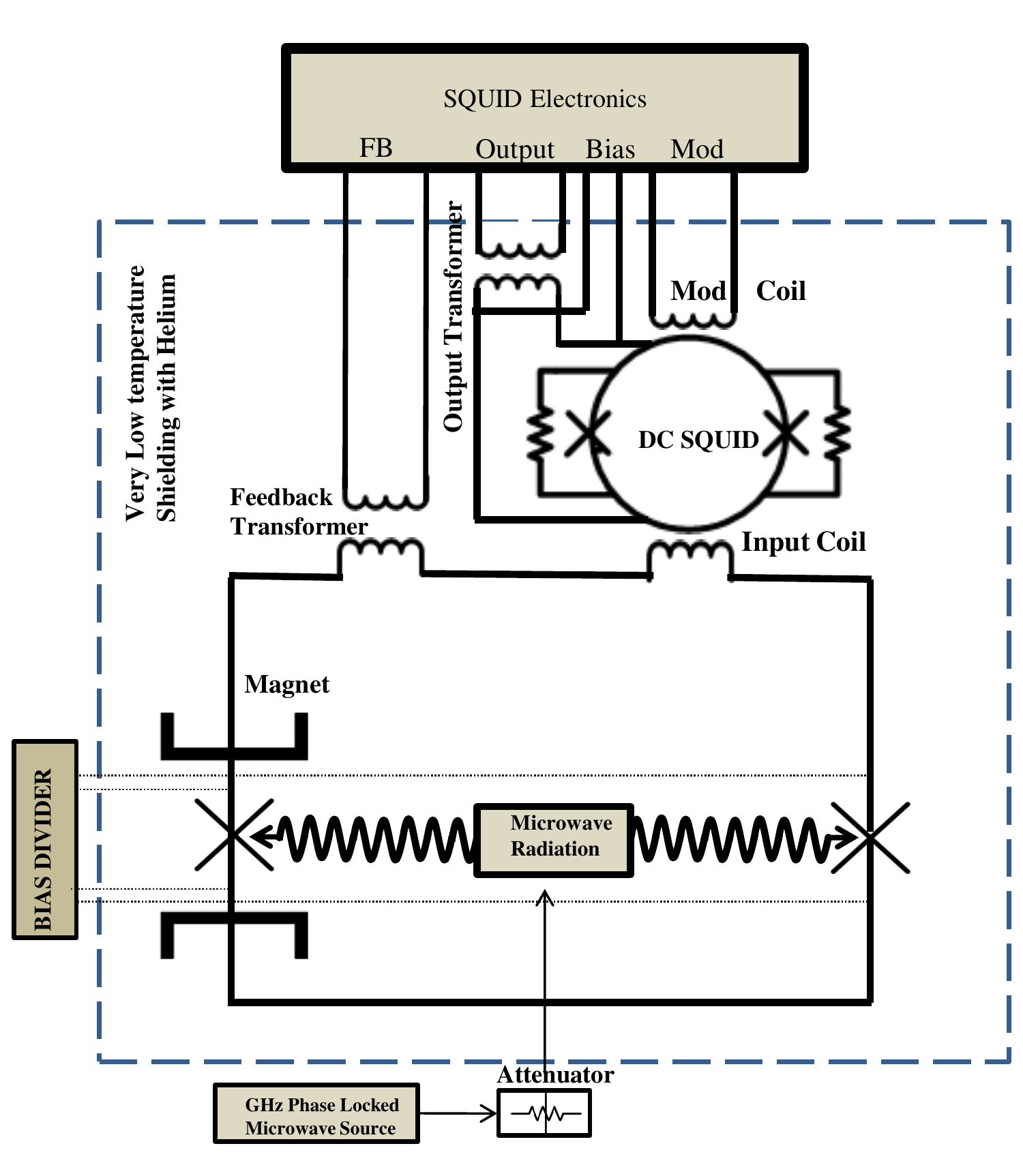}
  \caption{
The sketch  of proposed setup. Two Josephson junction has been located in superconducting loop. on/off magnet has been provided to influence left Josephson junction with strong magnetic field. Microwave radiates to both junctions and the different induced voltage makes a net current which increase by time linearly. the passed current from input coil provides magnetic field and flux. DC-SQUID setup has been designed to detect value of magnetic flux precisely. The main element should be located in liquid helium around 4 Kelvin for decreasing any source of noise.
 }
  \end{figure}
  \end{center}
Shortly after discovery of Josephson effects \cite{8}, many research groups planned to test the universality of this effect. As it has been proved, this effect is independent of type of junction with very high accuracy.
The special setup which J.Clarke proposed in early 1960\cite{4} approved this relation with accuracy up to $10^{-8}$ and further experiments, such as Tsai, Jain and Lukens in early 1980\cite{5} based on Clarke's experiment, improved the accuracy up to $10^{-19}$ by the technology of two decades ago.


This successful set up can inspire and help us to measure vacuum polarization of QED in presence of magnetic field. Our proposed set up has be shown in Fig.1. There is a junction loop which the left junction is located at local magnetic field and the other one in not affected by extra magnetic field. According to above arguments, the Josephson constant, $K_J:=2e/\hbar$ is different for this two  junctions. If we put the junctions on external radiation with frequency $\nu$ the relation of between voltage difference across the junction of left and right and frequency will given by 
\begin{equation}
V_{L,R}=n_s\nu {\Phi}_{0,L,R}= n_s \nu \frac{h}{2e_{L,R}}
\end{equation}
In order to obtaining the behavior of the main circuit, we write the relation between flux as following:
\begin{equation}
\frac{{\Phi}_{0,L}}{2\pi}{\theta}_L-\frac{{\Phi}_{0,R}}{2\pi}{\theta}_R+LI+{{\Phi}_{ext}}=m{\Phi}_0
\end{equation}
where ${\theta}_i$ is the junction phase. Variation  of the  above equation respect to time yields to
\begin{eqnarray}
L\frac{dI}{dt}&&=\Delta V=n\nu(\frac{1}{K_j^{L}}-\frac{1}{K_J^{R}})\\
&&\simeq V \frac{1}{45}\frac{\alpha}{\pi}(\frac{\hbar eB}{c^2m^2})^2 =V \times 10^{-24}(\frac{B}{0.1T})^2
\end{eqnarray}
this very simple equation shows that the current of the circuit grows linearly by time,
\begin{equation}
{I_{\text{Read-out}}} \propto \frac{e^4 \hbar}{c^5 {m_e}^4} t
\end{equation}
Estimating the created flux with current loop by SQUID can help us to measure the effect of vacuum polarization by calculating the slope  on linear line.


In the next section we discuses how to deal with tiny amount of voltage and current to have safe and precise measurement of this novel phenomena.

\section{Design of The Junctions}

\begin{center}
\begin{figure}[h]
  \includegraphics[height=80mm]{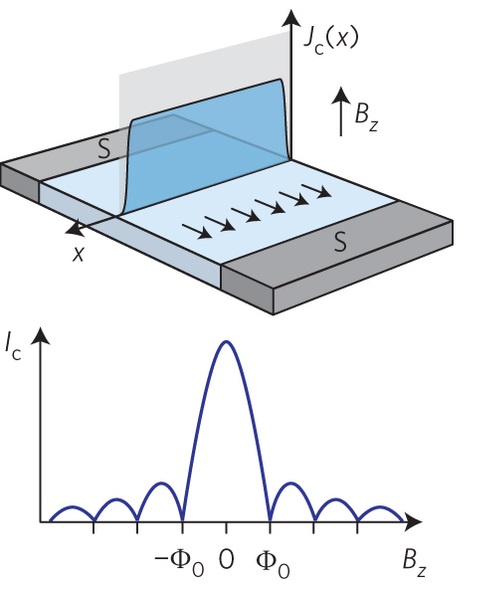}
  \caption{
  Fraunhofer pattern and a simple sketch of the Josephson junction and direction of the external magnetic field. The current density across the border junction changes due to external magnetic field, but that doesn't affect the calculate result because of the net current $\int J dx=I$ is important in Ac Josephson effect. The maximum current (critical current) that can pass through the junction is a function of magnetic field.  This upper bound is much higher that level of current we will dealing in this experiment for few hours running time.
  Picture taken from \cite{10}
 }
  \end{figure}
  \end{center}

The amount of  correction that we are looking for is related to projected magnetic field, $B$, quadratically.  On the other hand increasing the magnetic field deform the property of junction and we don't have emphases on increasing amount of magnetic field because it causes unexpected behavior from superconducting materials. However, the effect of correction on the output of circuit can increase by good junction design. for example array of Josephson junctions instead of single one can increase the net current linearly by number of series  junctions.



Selecting superconducting type II allow us to use magnetic field by mean value around $0.1 T$ safely.

 We can use a typical electromagnetic radiation with frequency of $
 f \sim 16 \text{GHz}$ in $10^{\text{th}}$ step, and this radiation provide suitable net voltage of
$ V \sim 300 \mu V$ across junction.

The inductance value of the junction loop depends on many factor, but    typical value can be considered  near to $L \sim 1nH$. By this parameters, current loop versus time due to vacuum polarization given by
\begin{equation*}
I(A) \sim 1.3 \ \  \text{fA}\times [\frac{(\frac{f}{16 \text{GHz}})(\frac{\# \text{of junction}}{1})(\frac{B}{0.1T})^2\times (\frac{T_c}{\text{Hour}})}{\frac{L}{1nH}}]
\end{equation*}

It's important to notice that existence of external magnetic  decrease critical current according to Josephson-Franhufer pattern \cite{3}.


If we consider the Josephson junction with barrier thickness $d$, width of strip line of each pair with $L$, and penetration depth with $\lambda$, then critical current will given by 
\begin{equation*}
I_c(B)\sim 14.7\ \ \mu \text{A} \times [\frac{\frac{I_{B=0}^{c}}{900\mu A}}{(\frac{B}{0.1T})(\frac{L}{1\mu \text{m}})(\frac{\lambda}{200 \text{nm}})}].
\end{equation*}



 This amount of critical current is  very high in compare with typical current we expect from doing experiment and  we need to run experiment consciously around one year to catch that current. Therefore we do not need to be worry about higher bound on current due to existence of external magnetic field.


\section{SQUID as a detector}

The Superconducting quantum interference device (SQUID)  is a very sensitive magnetometer which can be constructed by two or one Josephson junction. DC-SQUID is built from two Josephson junction on one conducting loop, and it can measure the magnetic field up to $5 \times 10^{-18}T$ and the low noise measurement of order Femto-Tesla is accessible in common research laboratories. The SQUID is one of the best detector of our small current in designed circuit.  We can joint the main superconducting loop to SQUID loop by two pickup coil, as it has illustrated in Fig.2.

The loop currents creates the magnetic flux that can estimated for circular loop with radius $R\sim 10 \text{cm}$ and $L \sim 1 \text{nH}$ as a following
\begin{equation*}
\Phi(T_c)= 1 \ \  m \Phi_0 \times (\frac{T_c}{\text{7 Hour}})(\frac{f}{483 \text{GHz}})(\frac{\# \text{of junc.}}{1})(\frac{B}{0.3T})^2
\end{equation*}
this amount of flux is in completely in feasible range of detection and it   increases linearly by time. Now days precision for magnetic flux is around $\mu \Phi_0$. Also the magnetic field inside the provided coil can be reach to $3 nT$ after around three hour experiment.  it's useful to compare this value with  the human brains magnetic field is of order $10^{-13} T$ and  magnetic field of heart which is  around $10^{-13}T$. Also The magnetometer of Gravity-Probe-B which constructed by SQUID, were sensitive to $10^{-18}T$, and  the effective threshold for  SQUIDs with current technology  is around $10^{-15}T$. 

\section{Noise Reduction }

In this section, we would like to show an advantage of signal processing and statistical analysis on the improvement of signal to noise ratio in our experiment. The key ingredient is that we have this ability to On and Off our observables and that profoundly helps us to remove many kinds of possible errors in our setup.


We can consider two kind of signal. The first one is the output of the circuit when external magnetic field  has been applied to junction and the second one is the readout with zero external magnetic filed,
\begin{equation}\label{twosignal}
\begin{split}
    \mathcal{X}_1&=\mathcal{S}_{B=\hat{B}}+\mathcal{N}_{1}\\
  \mathcal{X}_2&=\mathcal{S}_{B=0}+\mathcal{N}_{2'}:=\mathcal{N}_{2}
    \end{split}
\end{equation}

 Our expectation from $\mathcal{S}_{B}$, based on our theoretical calculations is a linear function in term of time and similar to any experiment in a real laboratory we should deal with noises were the sum of all kind of them had been denoted by $\mathcal{N}$ above. Modeling of noise is general depends on setup details but the major contribution of noises are thermal noises, and a stationary random process can model them. In practice, most of the random data that representing stationary physical phenomena are ergodic. This assumption and related facts help us to extract a useful date from observed results of in the limited number of experiments.

  If we measure a time series of stochastic process denoting by $\{ \mathcal{Y}(t)\}$ over specific period of time, $0 \leq t \leq T $ then the associated expectation value of that random variable define by 
 \begin{equation}
 \mathbb{E}[\mathcal{Y}]=\frac{1}{T}\int^{T}_{0}\mathcal{Y}(t)dt
 \end{equation}

 Auto-correlation and Cross-Correlation are  another useful quantities, which can be define  based on  two random variable $\mathcal{Y}$ and $\mathcal{X}$, 
 \begin{equation}
 \begin{split}
 \mathcal{R}_{\mathcal{Y}\mathcal{Y}}[\tau]&=\mathbb{E}[\mathcal{Y}(t)\mathcal{Y}(t+\tau)]\\
  \mathcal{R}_{\mathcal{X}\mathcal{Y}}[\tau]&=\mathbb{E}[\mathcal{X}(t)\mathcal{Y}(t+\tau)]
  \end{split}
 \end{equation}
 
Finally, the power spectral density of signal  define by Fourier transform of correlations are very useful quantities  as well:
 \begin{equation}\label{powerspec}
 \mathcal{P}_{\mathcal{X}\mathcal{Y}}[f]=\int_{-\infty}^{\infty} \mathcal{R}_{\mathcal{X}\mathcal{Y}}[\tau]e^{-j2\pi f \tau} d\tau
 \end{equation}
 
These definitions give us the great skills to extract our real signal from background noises. We now show how signal processing can remarkably increase the signal to noise ration in our proposed experiment.

According to Eq.\ref{twosignal}-\ref{powerspec},
\begin{equation}
\begin{split}
\mathcal{P}_{\mathcal{X}_1\mathcal{X}_1}&=\mathcal{P}_{\mathcal{S}\mathcal{S}}+\mathcal{P}_{\mathcal{N}_1\mathcal{N}_1}\\
\mathcal{P}_{\mathcal{X}_2\mathcal{X}_2}&=\mathcal{P}_{\mathcal{N}_2\mathcal{N}_2}\\
\mathcal{P}_{\mathcal{X}_1\mathcal{X}_2}&=\mathcal{P}_{\mathcal{S}\mathcal{S}}
\end{split}
\end{equation}

As we can see we can mind the power spectral density of our original signal from background noise by taking cross correlation. 


 It is  useful to calculate the Signal to noise ratio of our signals in this experiment. let's start with studying power spectrum and Fourier transform of correlations:

 If we represent the read out signal of the experiment by $\mathcal{S}(t)$, it is linear in time:
 \begin{equation}
     \mathcal{S}(t)=\Lambda t, \qquad 0 \leq t \leq T_c
 \end{equation}
 
 This implies that cross correlation of $\mathcal{S}(t)$ can given by
 
 
 
 \begin{equation}
 \begin{aligned}
 \mathcal{R}_{\mathcal{S}\mathcal{S}}[\tau]&= -\frac{\Lambda^2}{6}[2(T_c-\tau)^2(2T_c+\tau)\theta(T_c-\tau)\\
 &-(4T_c^3-6T_c^2\tau+\tau^3)\theta(2T_c-\tau)-\tau^3\theta(-\tau)]
 \end{aligned}
 \end{equation}
 
similarly, power spectrum density of signal will be 
 \begin{equation}
 \begin{aligned}
 \mathcal{P}_{\mathcal{S}\mathcal{S}}(f)&=4\int_{0}^{\infty}\mathcal{R}_{\mathcal{S}\mathcal{S}}[\tau]\cos(2\pi f \tau)d\tau\\
 &=\frac{\Lambda^2}{4\pi^4 f^4}( 1+(1-4f^2\pi^2 T_c^2)\cos(2\pi f T_c)\\
 &+4\pi f T_c(-\sin(2\pi f T_c)+\sin(4\pi f T_c)-2\cos(2\pi f T_c)),
 \end{aligned}
 \end{equation}
 hence
 \begin{equation}
 \mathcal{P}_{\mathcal{SS}}(f\rightarrow 0 )=T_c^4 \Lambda^2.
 \end{equation}

  The calculation of noise power depends on models that we consider for our noise. One of the most common noises in electrical circuits is Johnson–Nyquist noise. In Quantum Hall effect experiment this noise which relates to thermal fluctuations makes the situation hard for distinguishing physical phenomena from thermal noises. The spectral density of this noise given by:
 \begin{equation}
 \mathcal{P}_{\mathcal{NN}}=4k_{B}TR
 \end{equation}
 Where $T$ represents the temperature in Kelvin and $R$ is a resistor of an effective path of the circuit. For room temperature and resistance  around $1K\Omega$ we have
 \begin{equation}
 \sqrt{\mathcal{P}_{\mathcal{NN}}}\simeq \frac{4 nV}{\sqrt{Hz}}
 \end{equation}
 
 Now we can calculate the signl to noise ratio,
 \begin{equation}
 \frac{S}{N}=\frac{\sqrt{\mathcal{P}_{\mathcal{SS}(f\rightarrow 0)}}}{\sqrt{\mathcal{P}_{\mathcal{NN}}}}=\frac{T_c^2 \Lambda}{\sqrt{4k_B T R}}
 \end{equation}
 
 As we can see the SNR increases quadratically by a time interval of measurement and that the most promising feature of this proposed experiment to discover the fine structure of  nature.

Numerical calculation shows for typical value that we selected in previous sections we get  :
\begin{equation}
\text{SNR}( T_c) \approx 1.5 \times 10^{-7} (\frac{T_c}{\text{Sec}})^2=1.95 (\frac{T_c}{\text{Hour}})^2
\end{equation}
Which means after around 45 Min the singal power surmount the   noise power. After 10 hours, we will get $SNR \approx 195 $ or $3 \sigma$.

 \begin{center}
\begin{figure}[h]\label{001}
  \includegraphics[height=70mm]{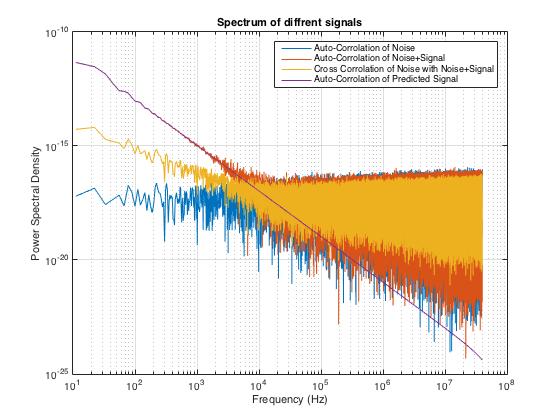}
  \caption{
 Simulated power spectral density of different Signals for run time $T=10$ hour. Recalling by definiton of  $\mathcal{X}_1 =\mathcal{S}_{B=\hat{B}}+\mathcal{N}_{1}$ and
  $\mathcal{X}_2=\mathcal{S}_{B=0}+\mathcal{N}_{2'}:=\mathcal{N}_{2}$, We can Simulate  the $\mathcal{P}(\mathcal{X}_1 \mathcal{X}_1), \mathcal{P}(\mathcal{X}_2 \mathcal{X}_2),\mathcal{P}(\mathcal{X}_2\mathcal{X}_1),\mathcal{P}(\mathcal{S} \mathcal{S}) $and$ \mathcal{P}(\mathcal{N}_1 \mathcal{N}_1)  $ by generating random noise and expected linear function.   increasing running time of experiment yields to  asymptotically merging  of  $\mathcal{P}(\mathcal{X}_1 \mathcal{X}_2)$ and $\mathcal{P}(\mathcal{X}_1\mathcal{X}_1)$  to $ \mathcal{P}(\mathcal{S}\mathcal{S})$(Purple line). 
 }
  \end{figure}
  \end{center}

\section{Technical notes}

Physical parameters that we work in this experiment are pretty hard to detect them in regular set ups and therefore precision in measurement and finding the sources of error is so crucial. Hopefully, in this experiment we have this ability to run the experiment with two modes, on and off an external magnetic field. Thus decreasing the outputs from these two methods statistically delete systematical errors. The only remaining concern is to find the noise source and making sure that signal to noise remains reasonable. Here we count some of them:
\begin{enumerate}
\item DC-bias can be anti-symmetric,  and It can have a drift by passing the time. Having symmetric current source with accuracy 1 in $10^6$ for measuring the symmetric properties of current sources help us to neglect this noise source possibly.
\item The frequency of microwave source should have no drift during the experiment, and the drift should be less than 100 Hz per day.
\item Attenuation of coaxial wire as waveguide should be constant during the measurement. This radiation drift can be a significant source of drift and error in this experiment also it is necessary that the output power should be constant
\item The most important issue that one should be careful in concection of SQUID magnetometer to the circuit. 
There is a thermoelectric voltage between the cryostat and the SQUID probe and the thermoelectric current flows through shell shielding of the circuit. This source of error  can significantly reduce by reducing the   distance between the cryostat and the SQUID probe close to the SQUID body.
\end{enumerate}

\section{Acknowledgment }

We thank Hessamadin Arfaei, Mohammad Amin, Jenny Hoffman, Mohammad Hafezi, Stephan Myer and Alexander Penin for helpful discussions. This work down in Superconductor Electronics Research Laboratory at Sharif University of Technology

\end{document}